\documentclass[]{aa}  

\usepackage[utf8]{inputenc}

\usepackage{graphicx}
\usepackage{txfonts}
\usepackage{hyperref}
\newcommand{\ham}{\mathcal{H}}
\newcommand{\hamsec}{\ham_{\mathrm{sec}}}
\newcommand{\hamsecGR}{\ham_{\mathrm{sec,\,GR}}}
\newcommand{\hamLL}{\ham_{\mathrm{LL}}}
\newcommand{\tsec}{t_{\mathrm{sec}}}
\newcommand{\tLyap}{t_{\mathrm{Lyap}}}

\begin{document}

\title{Addressing the statistical mechanics of planet orbits\\in the solar system}
%\subtitle{}
\author{Federico Mogavero}
\institute{Institut d’Astrophysique de Paris, 98 bis bd Arago, 75014 Paris, France\thanks{Sorbonne Universités, UPMC Paris 6 \& CNRS, UMR 7095.}\\
\email{mogavero@iap.fr}}

\date{Received ; accepted }

% \abstract{}{}{}{}{} 
% 5 {} token are mandatory
 
\abstract{The chaotic nature of planet dynamics in the solar system suggests the relevance of a statistical approach to planetary orbits. In such a statistical description, the time-dependent position and velocity of the planets are replaced by the probability density function (PDF) of their orbital elements. It is quite natural to set up this kind of approach in the framework of statistical mechanics. In the present paper I focus on the collisionless excitation of eccentricities and inclinations by gravitational interactions in a planetary system, the prototype of such a dynamics being the future planet trajectories in the solar system. I thus address the statistical mechanics of the planetary orbits in the solar system and try to reproduce the PDFs numerically constructed by Laskar (2008). I show that the microcanonical ensemble of the Laplace-Lagrange theory accurately reproduce the statistics of the giant planet orbits. To model the inner planets I then investigate the ansatz of equiprobability in the phase space constrained by the secular integrals of motion. The eccentricity and inclination PDFs of Earth and Venus are reproduced with no free parameters. Within the limitations of a stationary model, the predictions also show a reasonable agreement with Mars PDFs and that of Mercury inclination. The eccentricity of Mercury demands in contrast a deeper analysis. I finally revisit Laskar's random walk approach to the time dependence of the inner planet PDFs. Such a statistical theory could be combined with direct numerical simulations of planet trajectories in the context of planet formation, which is likely to be a chaotic process.}


\maketitle
%
%-------------------------------------------------------------------

\section{Introduction}
The chaotic dynamics of solar system planets \citep{laskar1989, sussman} raises the question of a statistical approach to planetary orbits. When chaos is significant, a single integration of the equations of motion is not representative of the entire possible dynamics. A description based on the probability density function (PDF) of the planet orbital elements, instead of their time-dependent position and velocity, is essentially more suitable. Such a statistical approach could be combined with usual direct numerical integrations, especially in the context of planet formation, which is likely to be a chaotic process \citep[e.g.,][]{hoffmann2017}.

The role of statistical mechanics in assessing the PDF of the planet orbital elements naturally emerges. Recently, the statistical mechanics of terrestrial planet formation has been addressed by \citet{tremaine}. In the spirit of a packed planetary 
systems hypothesis \citep{fang2013}, he advances the equiprobability of 
phase-space configurations verifying a certain criterion for the long-term stability of 
planetary systems. Within the limitations due to the sheared-sheet approximation 
and the restriction to the planar case, this ansatz allows to analytically compute 
any property derivable from the complete N-planet distribution function, such as the distributions of planet eccentricities and semimajor axis differences.
The predictions show an encouraging agreement with N-body simulations and
data from the \emph{Kepler} catalogue, while significant discordances arise in 
the case of radial-velocity data and solar system planets. The simplicity of Tremaine's ansatz is fundamental to describe at once the two main steps of the final 
giant-impact phase of terrestrial planet formation: the excitation of 
embryo eccentricities and inclinations through mutual gravitational interactions 
and the subsequent collisions and merging. However, one could ask for a more 
fundamental approach, such as one exploiting the integrals of motion \citep[e.g.,][]{Kocsis, touma2014, petrovich}, in line with the spirit of 
equilibrium statistical mechanics \citep{landau}. Such an approach can be actually set up by focusing on how eccentricities and inclinations are excited by gravitational interactions in a early, collisionless final phase of terrestrial planet formation, and postponing the analysis of collisions to subsequent treatments. As there is no fundamental physical difference between this problem and that of future planetary trajectories in the solar system \citep{laskar1996}, the latter can provide a benchmark to any statistical theory of planetary orbits to be applied to exoplanet systems. In the present study I address the problem of setting up the statistical mechanics of planetary trajectories in the solar system.

The statistical mechanics of gravitating systems is notoriously challenging \citep{padmanabhan}. Firstly, the system has to be confined in a spherical box to assure a bounded phase space, similarly to ideal gases. Then, the non-extensive nature of gravitational energy, due to the long-range character of Newtonian potential, breaks the equivalence between the microcanonical and canonical ensembles, which is typical of systems like neutral gases and plasmas. The long-range nature of gravity also prevents the existence of a thermodynamic limit. In addition to this, the short-range singularity of the gravitational potential requires to take into account the specific nature of small-scale interactions to guarantee the existence of microcanonical equilibrium states. Finally, the number $N$ of planets in a typical planetary system, even during the giant-impact phase of its formation process, is limited. Therefore, the $N \gg 1$ regime is generally inappropriate and so is the employment of a mean field approach\footnote{The lack of the $N \rightarrow + \infty$ limit could seem to question the usefulness of employing statistical mechanics. Actually, the foundation for such an approach relies on the chaotic nature of planet dynamics, independently of their number $N$.}. As pointed out by \citet{touma2014} mainly in the context of stellar systems, one effective way to construct the statistical mechanics of orbits in a planetary system is to average the planet motion over its fastest timescale, that is the orbital period. As a result of this averaging procedure, each planet is replaced by a massive ring following its Keplerian orbit, whose linear mass density is inversely proportional to the planet orbital velocity, a so-called Gaussian ring \citep{kondratyev}. In such a system, the rings interact with each other via secular gravitational interactions. Their eccentricities and inclinations relax by exchange of angular momentum, while their semi-major axes are constant, and so are their Keplerian energies \citep{rauch}. Even if there is no theorem assuring this secular dynamics approaches the actual planet motion, it can nevertheless represent the source of important results \citep{arnold}. Indeed, the first clear indications of a chaotic planetary motion in the solar system came from averaged equations of motion \citep{laskar1989}. Moreover, by using the same secular dynamics, \citet[from now on L08]{laskar2008} numerically computed the PDFs of eccentricity and inclination of the solar system planets. In the present paper I address the problem of reproducing these PDFs through a statistical mechanics approach.

The paper is structured as follows. In Sect.~\ref{secular_dynamics} I briefly recall how secular planet dynamics can be introduced in its simplest form. Then, in Sect.~\ref{laskar} I describe the PDFs of eccentricity and inclination of the solar system planets calculated numerically in L08. I introduce in Sect.~\ref{giant} the microcanonical ensemble of the Laplace-Lagrange theory to reproduce the statistics of the giant planet orbits. The ansatz of equiprobability in the phase space constrained by the secular integrals of motion is presented in Sect.~\ref{inner} for the inner planets. Finally, I revisit in Sect.~\ref{random_walk} the random walk ansatz of L08 to account for the time dependence of the inner planet PDFs. I conclude emphasizing the relevance of a statistical theory of planetary orbits in the context of planet formation.

%--------------------------------------------------------------------
\section{Secular planet dynamics}
\label{secular_dynamics}
Using standard notation \citep{morbidelli2002}, the Hamiltonian of the $N = 8$ solar system planets can be written as
\begin{equation}
\label{eq:hamiltonian}
\ham = \sum_{k = 1}^{N} \left( \frac{\vec{p}_k^2}{2 \mu_k} - G \frac{m_0 m_k}{r_k} \right) + \sum_{k = 1}^{N} \sum_{l=k+1}^{N} \left( \frac{\vec{p}_k \cdot \vec{p}_l}{m_0} - G \frac{m_k m_l}{| \vec{r}_k - \vec{r}_l |} \right)
\end{equation}
where heliocentric canonical variables $\vec{r}_k, \vec{p}_k$ are employed, $m_0$ is the Sun mass, $m_k$ the planet masses, $\mu_k = m_0 m_k/(m_0 + m_k)$ the reduced masses and $G$ the gravitational constant. As usual, the index $k$ lists the planets by increasing semi-major axis, from Mercury to Neptune. The first term on the right side of equation \eqref{eq:hamiltonian} is the Hamiltonian of the Keplerian motion, while the second one contains the gravitational interactions among planets. Therefore, it is worthwhile to introduce a set of action-angle variables for the Keplerian motion, e.g. the modified Delaunay variables \citep{morbidelli2002}:
\begin{align}
\label{eq:delaneuy}
  \begin{aligned}
   \Lambda_k &= \mu_k \sqrt{G(m_0 + m_k)a_k}   \\
   P_k &= \Lambda_k (1 - \sqrt{1-e_k^2})               \\
   Q_ k &= 2 \Lambda_k \sqrt{1-e_k^2} \sin^2(i_k/2)
  \end{aligned}
  &&
  \begin{aligned}
   \lambda_k &= M_k + \omega_k + \Omega_k   \\
   p_k &= - \omega_k - \Omega_k                        \\
   q_k &= - \Omega_k
  \end{aligned}
\end{align}
Keplerian orbital elements are used in these definitions: $a_k$ is the planet orbit semi-major axis, $e_k$ the eccentricity, $i_k$ the inclination, $M_k$ the mean anomaly, $\omega_k$ the argument of perihelion and $\Omega_k$ the longitude of node. As long as planets do not experience close encounters and do not lie near mean motion resonances, secular dynamics can be introduced by averaging the Hamiltonian \eqref{eq:hamiltonian} over the fastest motion timescales, i.e. the planet orbital periods. This corresponds to average the Hamiltonian over the mean anomalies $M_k$:
\begin{equation}
\label{eq:average}
\langle \ham \rangle = \frac{1}{(2 \pi)^N} \!\! \int_0^{2\pi} \!\!\! dM_1 \, \cdots \int_0^{2\pi} \!\!\! dM_N \, \, \ham = \ham_0 + \ham_{\text{sec}}
\end{equation}
where $\ham_0 = \sum_{k=1}^N -G(m_0+m_k)^2 \mu_k^3/(2 \Lambda_k^2) = \sum_{k=1}^N - G m_0 m_k/(2 a_k)$ is the total Keplerian energy and
\begin{equation}
\label{eq:secular}
\ham_{\text{sec}} =  - \sum_{k = 1}^{N} \sum_{l=k+1}^{N} \frac{G m_k m_l}{(2 \pi)^2} \int_0^{2 \pi} \!\!\! \int_0^{2 \pi} \! \frac{dM_k dM_l}{| \vec{r}_k - \vec{r}_l |} 
\end{equation}
are the orbit-averaged gravitational interactions between planets (the kinetic contribution in the second term on the right side of Eq. \eqref{eq:hamiltonian} averages out to zero over a Keplerian orbit). This averaging procedure corresponds to constructing a secular normal form to first order in planetary masses \citep{morbidelli2002}. The action variables $\Lambda_k$ are integrals of motion of the secular dynamics \citep{arnold}, and so are the semi-major axes $a_k$. As a consequence, $\ham_0$ is an additive constant which may be dropped from the Hamiltonian. The single-planet secular phase space is four-dimensional and compact, with the canonical volume element given by $dp_k dq_k dP_k dQ_k$. The secular contribution of general relativity is of primary importance for the long-term dynamics of the solar system inner planets (e.g., L08). The secular Hamiltonian accounting for the leading relativistic correction is given by
\begin{equation}
\label{eq:relativity}
\ham_{\text{sec}} =  - \sum_{k = 1}^{N} \! \sum_{l=k+1}^{N} \! \frac{G m_k m_l}{(2 \pi)^2} \! \int_0^{2 \pi} \!\!\! \int_0^{2 \pi} \! \frac{dM_k dM_l}{| \vec{r}_k - \vec{r}_l |}  - \sum_{k = 1}^{N} \frac{3 G^2 m_0^2 m_k}{c^2 a_k^2 \sqrt{1-e_k^2}}
\end{equation}
where $c$ is the speed of light. A short derivation of the general relativistic contribution is presented in the Appendix~\ref{secular_hamiltonian} for future reference.

When eccentricities and inclinations are sufficiently small, $e_k, \sin(i_k/2) \ll 1$, it is valuable to develop the secular Hamiltonian in a power series of these variables. Neglecting terms of the fourth order and smaller, one obtains the Laplace-Lagrange (LL) linear secular theory:
\begin{equation}
\label{eq:quad_hamiltonian}
\hamLL = - \frac{ \vec{x}^{\rm T} \tens{A} \vec{x} + \vec{y}^{\rm T} \tens{A} \vec{y} + \vec{v}^{\rm T} \tens{B} \vec{v} + \vec{z}^{\rm T} \tens{B} \vec{z} }{2}
\end{equation}
where I have introduced the Poincar\'{e} canonical variables,
\begin{align}
\label{eq:poincare_variables}
  \begin{aligned}
   x_k &= \sqrt{\Lambda_k} e_k \sin p_k   \\
   y_k &= \sqrt{\Lambda_k} e_k \cos p_k
  \end{aligned}
  &&
  \begin{aligned}
   v_k &= 2 \sqrt{\Lambda_k} \sin(i_k/2) \sin q_k   \\
   z_k &= 2 \sqrt{\Lambda_k} \sin(i_k/2) \cos q_k
  \end{aligned}
\end{align}
and, for instance, $\vec{x}^{\rm T} = (x_1, \dots, x_N)$. The elements of the matrices $\tens{A}$ and $\tens{B}$ are given in the Appendix \ref{matrices}. These matrices are real and symmetric and can therefore be diagonalized through orthogonal matrices $\tens{O_A}$ and $\tens{O_B}$,
\begin{equation}
\label{eq:diagonalization}
\tens{A} = \tens{O_A D_A O_A^{\/ \rm T}}, \quad \tens{B} = \tens{O_B D_B O_B^{\/ \rm T}}
\end{equation}
The elements of the diagonal matrices $\tens{D_A}$ and $\tens{D_B}$ are the eigenvalues of $\tens{A}$ and $\tens{B}$, respectively, and the columns of $\tens{O_A}$ and $\tens{O_B}$ are their normalized eigenvectors. New action-angle coordinates $\vec{P'}, \vec{Q'}, \vec{p'}, \vec{q'}$ can be introduced by combining the following canonical transformations:
\begin{align}
\label{eq:new_poincare_variables}
  \begin{aligned}
   \vec{x} &= \tens{O_A} \vec{x'} \\
   \vec{y} &= \tens{O_A} \vec{y'}
  \end{aligned}
  &&
  \begin{aligned}
   \vec{v} &= \tens{O_B} \vec{v'} \\
   \vec{z} &= \tens{O_B} \vec{z'}
  \end{aligned}
\end{align}
\begin{align}
\label{eq:new_delaunay}
  \begin{aligned}
   x'_k &= \sqrt{2 P'_k} \sin p'_k   \\
   y'_k &= \sqrt{2 P'_k} \cos p'_k
  \end{aligned}
  &&
  \begin{aligned}
   v'_k &= \sqrt{2 Q'_k} \sin q'_k   \\
   z'_k &= \sqrt{2 Q'_k} \cos q'_k
  \end{aligned}
\end{align}
Employing the new variables, the LL Hamiltonian becomes:
\begin{equation}
\label{eq:quad_hamiltonian_2}
\hamLL = - \sum_{k=1}^N \left( g_k P'_k + s_k Q'_k \right)
\end{equation}
where $g_k$ and $s_k$ are the eigenvalues of the matrices $\tens{A}$ and $\tens{B}$, respectively. According to the corresponding Hamilton equations, the actions $P'_k$ and $Q'_k$ are integrals of motion, while the angles $p'_k$ and $q'_k$ change linearly with time at constant frequencies $-g_k$ and $-s_k$, respectively. Therefore, the time dependence of the Poincar\'{e} variables $x_k, y_k, v_k, z_k$ turns out to be the superposition of $N$ harmonics with different frequencies and amplitudes. In the present study I refer to each of these harmonics as a LL mode and I follow the conventional ordering of the frequencies $g_k$ and $s_k$ \citep[Table 7.1]{morbidelli2002}.

\subsection{Integrals of motion}
\label{constants}
Generally speaking, the dynamics of a planetary system obeys to the conservation of mechanical energy, momentum and angular momentum. The secular dynamics introduced above automatically verifies the conservation of momentum: as Keplerian orbits are closed, the average of planet momenta over mean anomalies identically vanishes. Therefore, the dynamics resulting from the secular Hamiltonian \eqref{eq:relativity} must conserve energy and angular momentum. This can be easily shown to be the case in the LL dynamics \citep[e.g.,][]{Kocsis}. As formula \eqref{eq:quad_hamiltonian_2} shows, the LL Hamiltonian is a function of the action variables only. These being integrals of motion, the conservation of energy follows. The angular momentum content of a planetary system can be described by the following quantities:
\begin{equation}
\label{eq:angular_momentum}
\begin{aligned}
L_x &= \sum_{k=1}^N \Lambda_k \sqrt{1-e_k^2} \sin i_k \sin \Omega_k    \\
L_y &= - \sum_{k=1}^N \Lambda_k \sqrt{1-e_k^2} \sin i_k \cos \Omega_k    \\
C &= \sum_{k=1}^N \Lambda_k \left( 1- \sqrt{1-e_k^2} \cos i_k \right)
\end{aligned}
\end{equation}
where $x, y, z$ are the Cartesian coordinates related to the definition of the Keplerian elements $i_k, \omega_k$ and $\Omega_k$ \citep{morbidelli2002}\footnote{Typically, the $x\,y$\nobreakdash-reference plane is chosen to be perpendicular to the total angular momentum vector. However, in this study I choose the ecliptic plane as the reference one, since this is the choice made in L08.}. $L_x$ and $L_y$ are the components of the angular moment lying on the $x\,y$\nobreakdash-reference plane, while $C$ is the angular momentum deficit (AMD) \citep{laskar1997, laskar2000}, the difference between the angular momentum that planets would have on coplanar, circular orbits and $L_z$, the component of the total angular momentum perpendicular to the reference plane. The AMD can be expressed as
\begin{equation}
\label{eq:deficit}
C = \sum_{k=1}^N P_k + Q_k = \sum_{k=1}^N P'_k + Q'_k
\end{equation}
where I have used the definitions \eqref{eq:delaneuy} and the fact that the matrices $\tens{O_A}$ and $\tens{O_B}$, appearing in the transformation \eqref{eq:new_poincare_variables}, are orthogonal. As the action variables $\vec{P'}$ and $\vec{Q'}$ are conserved in the LL dynamics, the AMD is also a conserved quantity. Conservation of $L_x$ and $L_y$ derives from the properties of the matrix $\tens{B}$. As shown in the Appendix~\ref{matrices}, one of the eigenvalues $s_k$ is zero and $\left. \left( \!\! \sqrt{\Lambda_1}, \dots, \! \sqrt{\Lambda_N} \right) \, \right/ \! \! \sqrt{\sum_k\Lambda_k}$ is, up to a sign, the corresponding normalized eigenvector. By convention \citep[Table 7.1]{morbidelli2002}, the null eigenvalue is chosen to be $s_5$. Moreover, neglecting terms of order three in eccentricities and inclinations, one has
\begin{equation}
\label{eq:angular_momentum_Laplace_Lagrange}
\begin{aligned}
L_x &= - \sum_{k=1}^N 2 \sqrt{\Lambda_k} v_k   \\
L_y &= - \sum_{k=1}^N 2 \sqrt{\Lambda_k} z_k 
\end{aligned}
\end{equation}
All this implies that in the LL dynamics the following identities are verified:
\begin{equation}
\label{eq:convervation_Lx_Ly}
\begin{aligned}
v'_5 = \sum_{k=1}^N \tens{O_B^{\/ \rm T}}_{5k} v_k = - \frac{L_x}{2 \sqrt{\sum_{k=1}^N \Lambda_k}} \\
z'_5 = \sum_{k=1}^N \tens{O_B^{\/ \rm T}}_{5k} z_k = - \frac{L_y}{2 \sqrt{\sum_{k=1}^N \Lambda_k}}
\end{aligned}
\end{equation}
where I have used the fact that $\tens{O_B}$ is orthogonal, $\tens{O_B^{\/ -1}} = \tens{O_B^{\/ \rm T}}$, and its columns are the normalized eigenvectors of $\tens{B}$. As the frequency $s_5$ is zero, $q'_5$ does not change with time. Therefore, $v'_5$ and $z'_5$ are integrals of motion and so are $L_x$ and $L_y$. 

\begin{figure*}
\centering
\includegraphics[width=17cm]{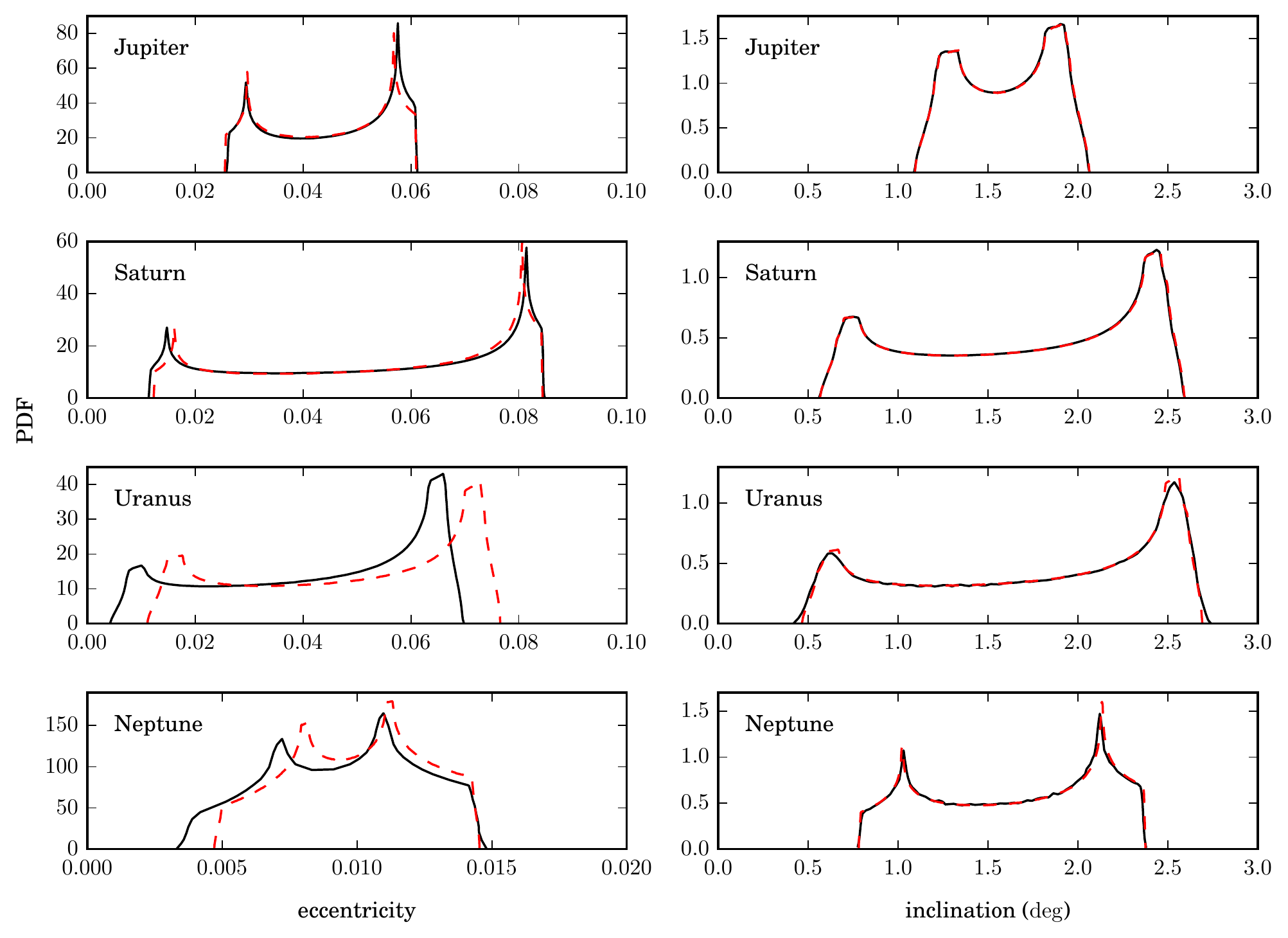}
\caption{The PDF of the eccentricity and inclination of the solar system giant planets. The PDFs of L08 are plotted in solid line, while those predicted by the microcanonical distribution function \eqref{eq:microcanonical_giant} are shown in dashed line. The inclinations are computed with respect to the J2000 ecliptic.}
\label{figure1}
\end{figure*}

\section{The PDFs of \citet{laskar2008}}
\label{laskar}
\begin{figure*}
\centering
\includegraphics[width=17cm]{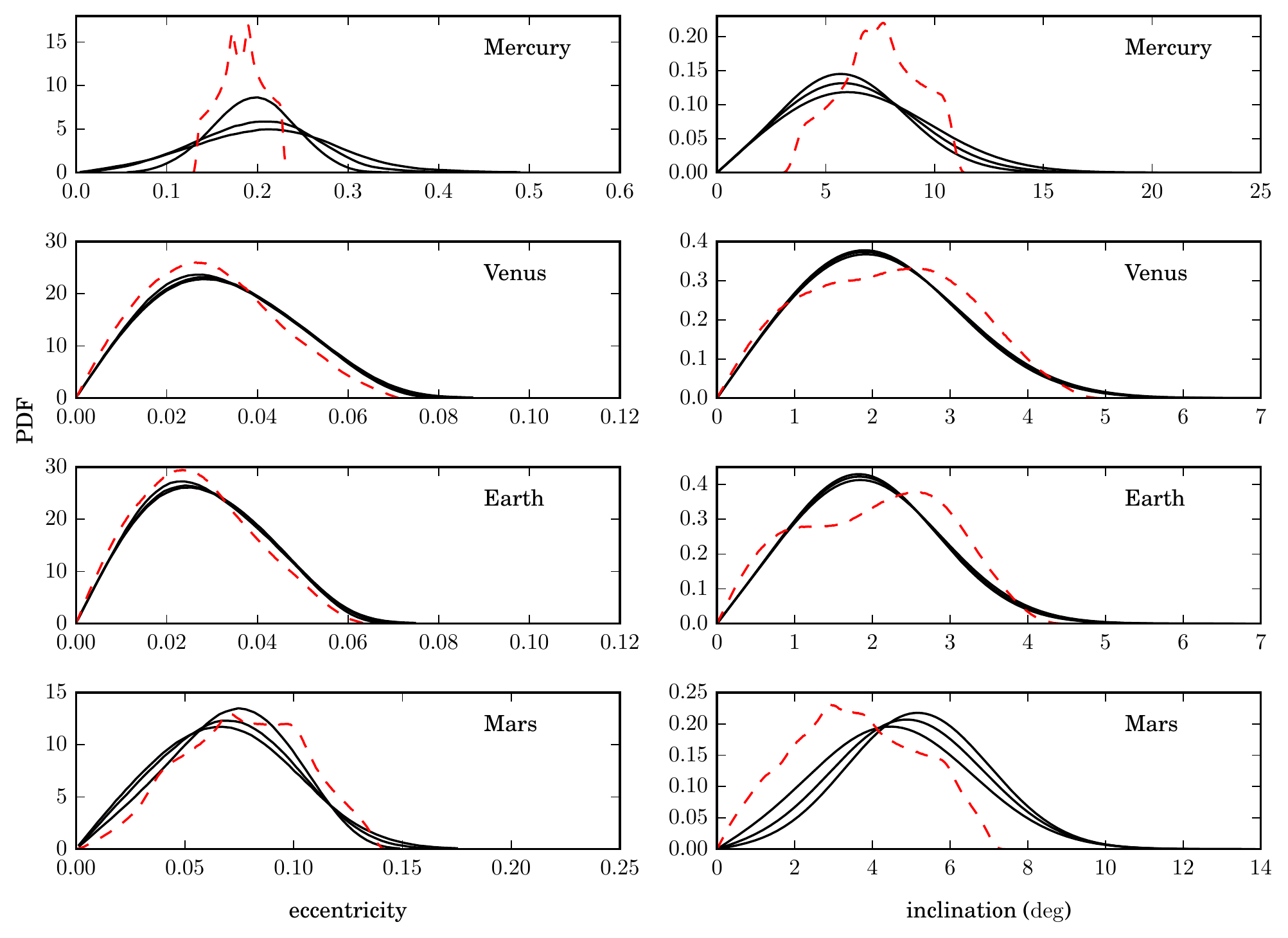}
\caption{The PDF of the eccentricity and inclination of the solar system inner planets. The PDFs of L08 are plotted in solid line at three different times, $t =$ 500 Myr, 2.5 Gyr and 5 Gyr, to illustrate chaotic diffusion (the PDF broadens and its peak decreases as time increases. The inclination PDFs are not the original ones but Laskar's fits to them, see Footnote \ref{footnotePDF} for explanations). The PDFs predicted by the microcanonical distribution function \eqref{eq:microcanonical_giant} are shown in dashed line. The inclinations are computed with respect to the J2000 ecliptic.}
\label{figure2}
\end{figure*}

In L08 Laskar performs a statistical analysis of the future planet orbits in the solar system. He employs 1001 integrations of secular equations over the 5 Gyr of the Sun's remaining lifetime before its red giant phase. These integrations differ for the initial conditions, which are obtained with small variations of the initial Poincar\'{e} variables of the VSOP82 solution \citep{bretagnon1982}. The total integration time is divided in 250 Myr intervals and statistics are performed over each interval recording the state of the 1001 solutions with a 1 kyr timestep. Normalized PDFs of planet eccentricities and inclinations are then estimated. The PDFs of the giant planets are plotted in Fig. \ref{figure1}, while in Fig. \ref{figure2} are shown those of the inner ones\footnote{\label{footnotePDF}Unfortunately, the data behind the L08 PDFs are not currently available (Laskar, private communication). Therefore, I have extracted these curves from the original plots through WebPlotDigitizer \citep{webplot}. This works fine for the giant planets PDFs and the eccentricity PDFs of the inner ones, as Laskar plotted them separately at three different times. Because of chaotic diffusion, such an extraction is infeasible for the inclination PDFs of the inner planets, which Laskar plotted at several different times on the top of each other. In this case I use as a reference Laskar's fits instead of the original numerical PDFs, even if some differences exists between these curves, especially in the tails.}. The dynamics considered by Laskar is more accurate than the one I introduced in Sect. \ref{secular_dynamics}, as Laskar's secular Hamiltonian contains terms of order two in planetary masses and six in eccentricities and inclinations. This is why he conjectures that, even though his PDFs are obtained through secular equation, they should be nevertheless close to those arising from the full, non-averaged, dynamics. While analysing these PDFs, what strikes Laskar is the very different shape between the giant planets curves and those of the inner planets. The PDFs of the outer planets are characterized by two peaks and restricted to a certain range of eccentricities and inclinations. In contrast, those of the inner planets have zero value at $e = 0$ and $i = 0$, a single peak and a continuous decaying at large eccentricities and inclinations. On the basis of these differences, Laskar takes two different approaches in explaining his numerical PDFs. By means of a frequency analysis applied to the outer planet motion, he shows that the PDF of a quasiperiodic approximation of their dynamics can reproduce very accurately the numerical PDFs. On the other hand, to take into account the significant chaotic diffusion existing in the statistics of the inner planets, Laskar fits their PDFs through a Rice distribution \citep{rice1945}. Indeed, he assumes that, because of chaotic diffusion, the Poincar\'{e} variables \eqref{eq:poincare_variables} become practically independently Gaussian distributed over long timescales, with non-zero mean and a variance that increases linearly with time, like in a diffusion process.
\begin{figure*}
\centering
\includegraphics[width=17cm]{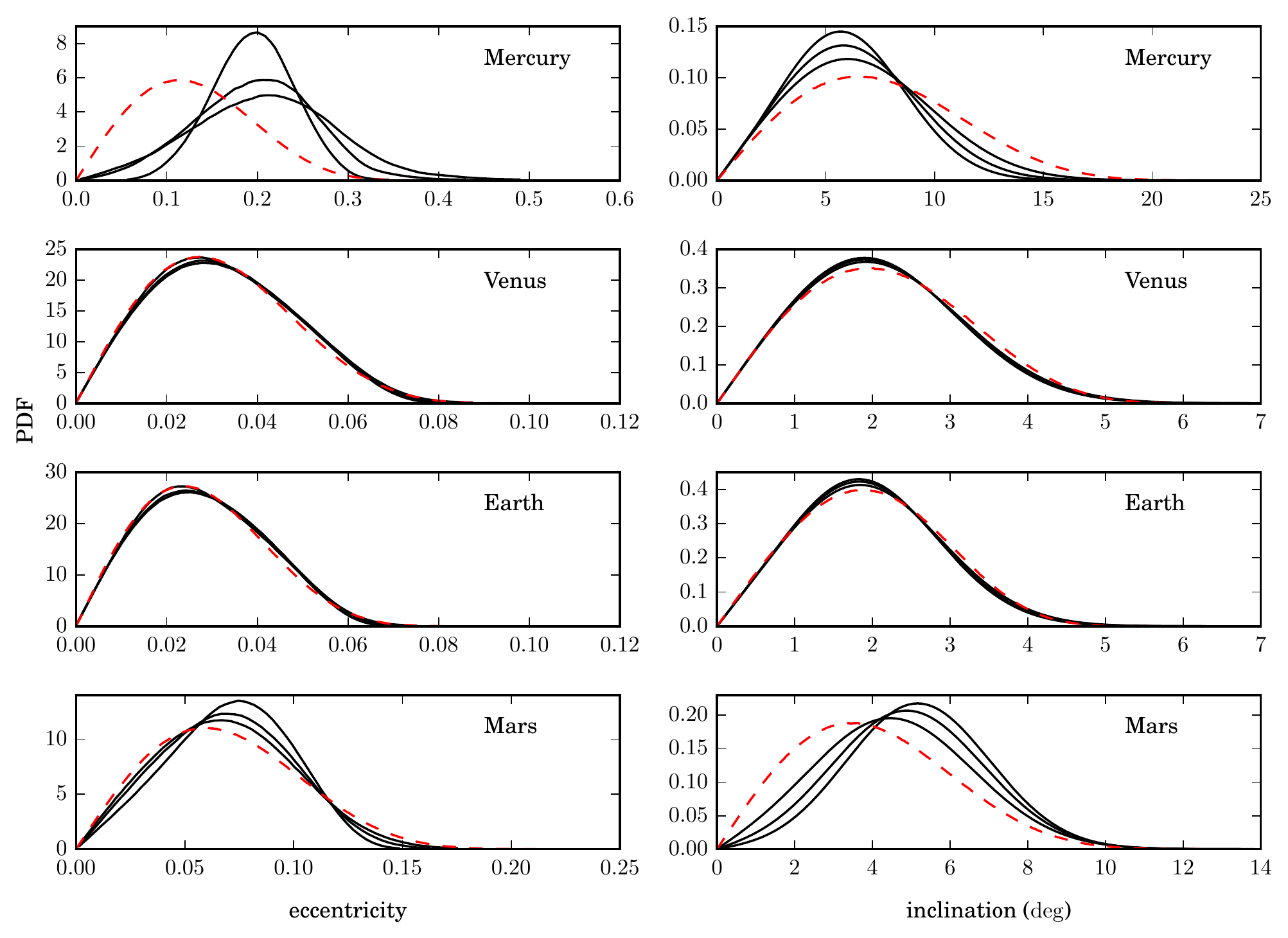}
\caption{The PDF of the eccentricity and inclination of the solar system inner planets. The same as Fig.~\ref{figure2}, except that the dashed line represents here the prediction of the ansatz~\eqref{eq:ergodic_ansatz}.}
\label{figure3}
\end{figure*}
\section{Giant planets}
\label{giant}
According to secular dynamics, the orbital motion of the giant planets is well approximated by a quasiperiodic time function (L08). Indeed, chaotic diffusion is very limited for the outer planets, their PDFs virtually do not change with time. A certain chaotic behaviour arises from the full equations of motion \citep{sussman}, but according L08 this is irrelevant to the overall structure of the PDFs. Taking advantage of this regularity of the giant planets, it is straightforward to predict the PDF of their Keplerian orbital elements. Neglecting chaotic diffusion, the most basic quasiperiodic dynamics is that resulting from the LL Hamiltonian, Eq.~\eqref{eq:quad_hamiltonian_2}. Such a motion is ergodic and the asymptotic stationary probability density is the microcanonical one \citep{arnold}. It is straightforward to write down the corresponding PDF employing the integrals of motion~\citep{landau}:
\begin{equation}
\label{eq:microcanonical_giant}
\rho(\vec{P'}, \vec{Q'}, \vec{p'}, \vec{q'}) = \frac{\delta(q'_5 - \overline{q'_{5}})}{(2 \pi)^{2N-1}} \prod_{k=1}^N \delta(P'_k - \overline{P'_{k}}) \, \delta(Q'_k - \overline{Q'_{k}})
\end{equation}
where $\delta$ stands for the Dirac delta and the bar indicates the initial value of the corresponding variable\footnote{In the present study these initial values are chosen to match the initial conditions of the VSOP82 solution \citep[Tables 4 and 5]{bretagnon1982}, since they are the ones employed in L08.}. This PDF reflects the conservation of the action variables $\vec{P'}, \vec{Q'}$ in the LL dynamics, while the angle variables are uniformly distributed over the ($2N$$-$$1$)-dimensional torus $\mathbf{T}^{2N-1}$ (the angle $q'_5$ is conserved because the angular momentum is, see Sect.~\ref{constants}). An integral formula for the PDFs of eccentricity and inclination can be obtained \citep{mardia1972},
\begin{equation}
\label{eq:microcanonical_giant_formula}
\rho(e_k) = e_k \int_0^{+ \infty} du \,\, u \, J_0(e_k u) \prod_{n = 1}^N J_0(\gamma_{kn} u)
\end{equation}
where $J_0$ is the Bessel function of the first kind and order zero and $\gamma_{kn} = \tens{O_A}_{kn} (2\overline{P'_n}/\Lambda_k)^{1/2}$. A similar formula holds for the inclination $i_k$. Otherwise, these PDFs can be very rapidly estimated by direct sampling of Eq.~\eqref{eq:microcanonical_giant} and using transformations~\eqref{eq:new_delaunay}, \eqref{eq:new_poincare_variables} and \eqref{eq:poincare_variables}. The distribution function \eqref{eq:microcanonical_giant} is time-independent and describes the statistics of planet orbits over timescales much longer than the timescale of the LL dynamics (dozen of kyr to few Myr for the solar system planets). Since Laskar's PDFs are constructed over 250-Myr intervals, one can expect the statistics of the giant planets resulting from Eq.~\eqref{eq:microcanonical_giant} to agree with that of L08. Therefore, I plot in Fig. \ref{figure1} the PDFs predicted by the microcanonical distribution \eqref{eq:microcanonical_giant}, along with Laskar's PDFs. The agreement between the two curves is indeed very satisfactory for the inclination PDFs, with a very good matching for Jupiter and Saturn and minor differences in the case of Uranus and Neptune. The agreement is also good for the eccentricity PDFs of Jupiter and Saturn, while some major discordances arise in the eccentricity PDFs of Uranus and Neptune. Even though the shape of the PDFs is correctly reproduced by the microcanonical density \eqref{eq:microcanonical_giant} in both cases, the endpoints of the eccentricity interval over which the PDFs vary are somewhat different from L08. However, the reason for such a discrepancy is clear. The LL Hamiltonian \eqref{eq:quad_hamiltonian_2} employed to derive the microcanonical distribution \eqref{eq:microcanonical_giant} is valid up to the third order in eccentricities and inclinations. This regime is meaningful for the giant planets of the solar system, whose eccentricities and inclinations are quite moderate. However, terms of the fourth order in $e$ and $i$ in the Hamiltonian \eqref{eq:relativity} yield non-linearities in the corresponding Hamilton equations and produce high-order harmonics in the eccentricity and inclination time dependence\footnote{These higher order terms also affect amplitude and frequency of the LL modes. It is worthwhile to remember that the secular Hamiltonian \eqref{eq:relativity} is itself valid to the first order in planetary masses. Contributions from higher order terms in $m_k$ slightly adjust amplitude and frequency of the harmonics.}. The amplitude of these additional harmonics is generally much smaller than that of the LL modes, but some of them can nevertheless contribute to a significant level. \citet[Tables 12 and 13]{bretagnon1974} reports the amplitudes of these high-order harmonics. From Bretagnon's tables it is clear that their contribution is bigger for the eccentricities than the inclinations and for Uranus and Neptune than Jupiter and Saturn. This is in agreement with Fig. \ref{figure1}. Moreover, the amplitudes of the higher order harmonics are roughly in accord with the size of the discrepancies in Fig. \ref{figure1}. Generally speaking, the agreement between the predictions of the microcanonical distribution \eqref{eq:microcanonical_giant} and Laskar's PDFs is very satisfactory when one realizes the simplicity of the LL approximation. Moreover, a better agreement can be obtained extending the analysis of Sect. \ref{secular_dynamics} to higher order terms in eccentricities, inclinations and masses. In principle, this could be done by means of a perturbation approach to the Hamiltonian \eqref{eq:hamiltonian} based on successive quasi-identity canonical transformations and normal forms \citep{morbidelli2002}.

\section{Inner planets}
\label{inner}
The analysis of the giant planet orbits is based on the assumption that the effect of chaos on their motion is largely negligible. This is not the case for the inner planets. \citet{laskar1994} already showed that Mercury and Mars orbits are affected by a significant chaotic diffusion over 5 Gyr, while this diffusion is moderate for Venus and the Earth. A first interesting step in analysing the structure of Laskar's PDFs of the inner planets is to consider what the distribution function \eqref{eq:microcanonical_giant} predicts for their eccentricity and inclination. This is illustrated in Fig. \ref{figure2}, where I also plot the corresponding L08 PDFs at three different times, $t =$ 500 Myr, 2.5 Gyr and 5 Gyr, to illustrate chaotic diffusion. As one might expect, there is no agreement between the two curves. Nevertheless, it is interesting to note that, contrary to what could emerge from Laskar's analysis in L08, the difference in shape between the giant and inner planet PDFs is not fundamental. Fig. \ref{figure2} shows that Eq. \eqref{eq:microcanonical_giant} based on the LL dynamics is a priori able to produce the general shape of the inner planet PDFs, i.e. the zero value and positive linear slope at $e = 0$ and $i=0$, and the continuous decaying at large eccentricities and inclinations. This is particularly evident in the Venus and Earth eccentricity curves and is due to the strong linear coupling existing between several LL modes in the Earth and Venus dynamics. Analogue strong couplings are not present in the LL dynamics of the giant planets. This explains the different shape of their PDFs, with two peaks and a zero probability outside a certain range (Fig. \ref{figure1}). This is also the case for Mercury, what justifies the strong differences between the corresponding Laskar's PDFs and Eq. \eqref{eq:microcanonical_giant} in Fig.~\ref{figure2}.

\subsection{The ansatz of equiprobability of phase-space states}
Since the effect of chaos is relevant for the long-term dynamics of the inner planets, I search for a more pertinent statistical description than one based on quasiperiodic motion. The chaotic nature of planet dynamics allows for sharing of angular momentum between the LL action variables, which would otherwise be constant. This chaotic diffusion of angular momentum is already mentioned in \citet{laskar1996} and has been recently illustrated in \citet{wu2011}. The simplest statistical approach is assuming that, by consequence, the motion of the inner planets uniformly visits the subset of the phase space defined by the conservation of angular momentum $\vec{L}$ and energy $\hamsec$. Clearly, the fact that chaotic diffusion finally lead to such an uniform phase-space measure cannot be strictly true, as the inner planet PDFs depend on time. More probably, an uniform exploration could indeed occur adiabatically on a smaller subset of the phase space than that allowed by the conservation of angular momentum and energy only. Such a domain could be in principle determined by studying the higher order terms in the Hamiltonian \eqref{eq:relativity}. However, the hypothesis of equiprobability in the phase-space is still the best estimate that one can make without a deeper analysis of the dynamics \citep{jaynes1957}. It so simple that it is worthwhile to investigate its predictions before undertaking more involved studies. Indeed, the idea of equipartition between LL modes has already been considered from a dynamical perspective by \citet{wu2011}. Moreover, even though it leads to a stationary distribution function, such an ansatz is nevertheless quite appropriate to Earth and Venus, since their PDFs diffuse very slowly over time, and can be still useful to understand the main factors contributing to the overall structure of Mercury and Mars PDFs.

\begin{figure}
\centering
\resizebox{\hsize}{!}{\includegraphics{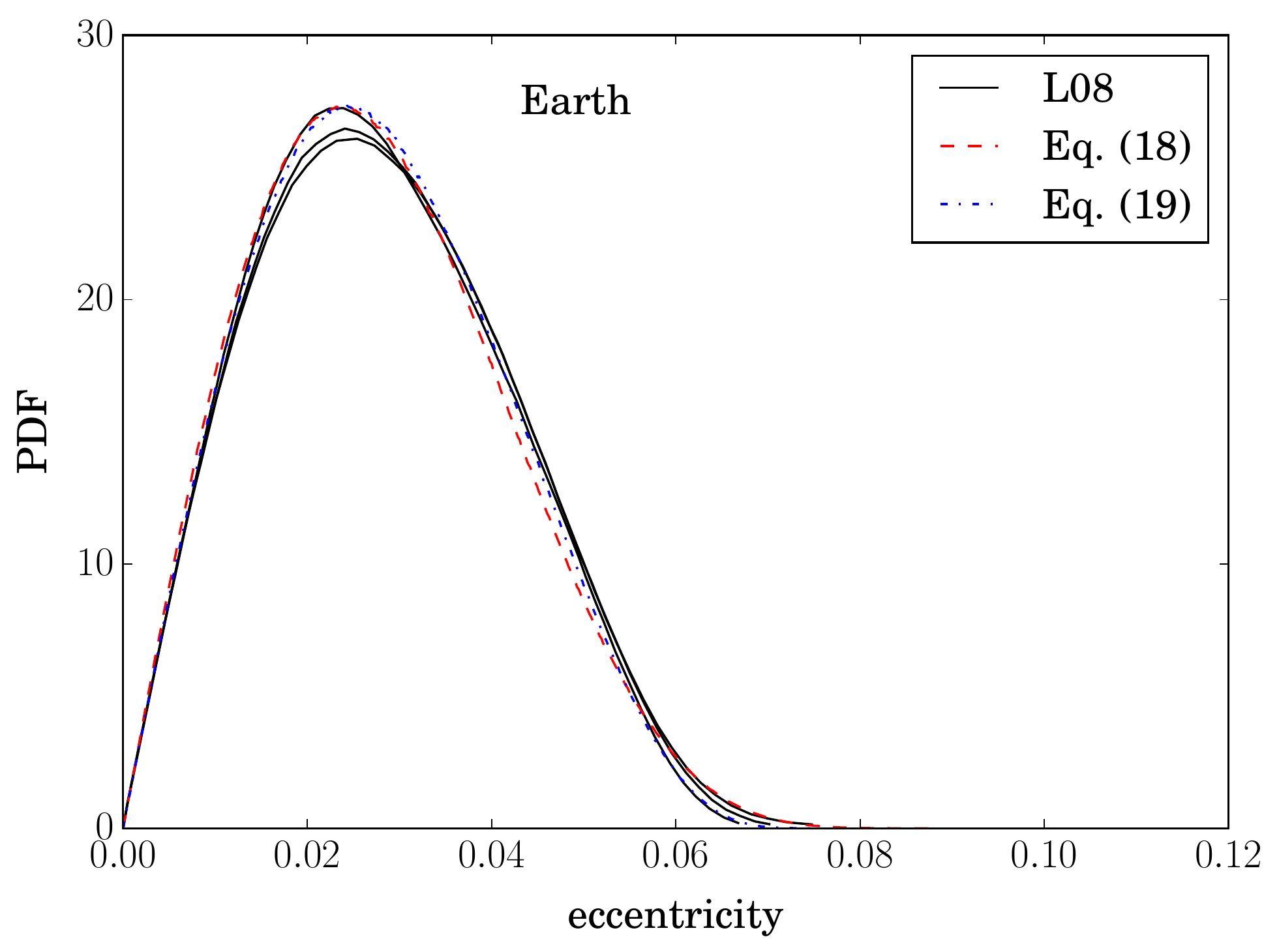}}
\caption{PDF of Earth eccentricity. The same as Fig.~\ref{figure3} with the addition of the prediction of Eq.~\eqref{eq:complete_ergodic_ansatz} in dash-dot line.}
\label{figure4}
\end{figure}

Since the giant planets are well described by the microcanonical distribution \eqref{eq:microcanonical_giant}, in which all the LL action variables are fixed, it is clear that an equiprobability simply based on conservation of the total angular momentum and energy of all the planets would not produce reasonable results. To achieve a global description of giant and inner planets at once, one needs to statistically disconnect them. To this end, I note that the total angular momentum of the inner planets is quite independent of that of the giant planets. More precisely, \citet{laskar1997} has shown that the flux of angular momentum deficit (AMD) from the large reservoir of the outer planets to the inner ones is moderate over 5 Gyr. As a first approximation, one can therefore assume that the total AMD of the inner planets is constant. Moreover, from Eq.~\eqref{eq:deficit} one can interpret the action variables $P'_k$ and $Q'_k$ as the AMD content in the corresponding LL mode. These considerations suggest to fix the action variables corresponding to the modes $k \in \{5, 6, 7, 8\}$, as they are the only ones which are relevant for the dynamics of the outer planets\footnote{As stated in Sect.~\ref{secular_dynamics}, I adopt the conventional ordering of the LL modes \citep[Table 7.1]{morbidelli2002}.}.
\begin{figure}
\centering
\resizebox{\hsize}{!}{\includegraphics{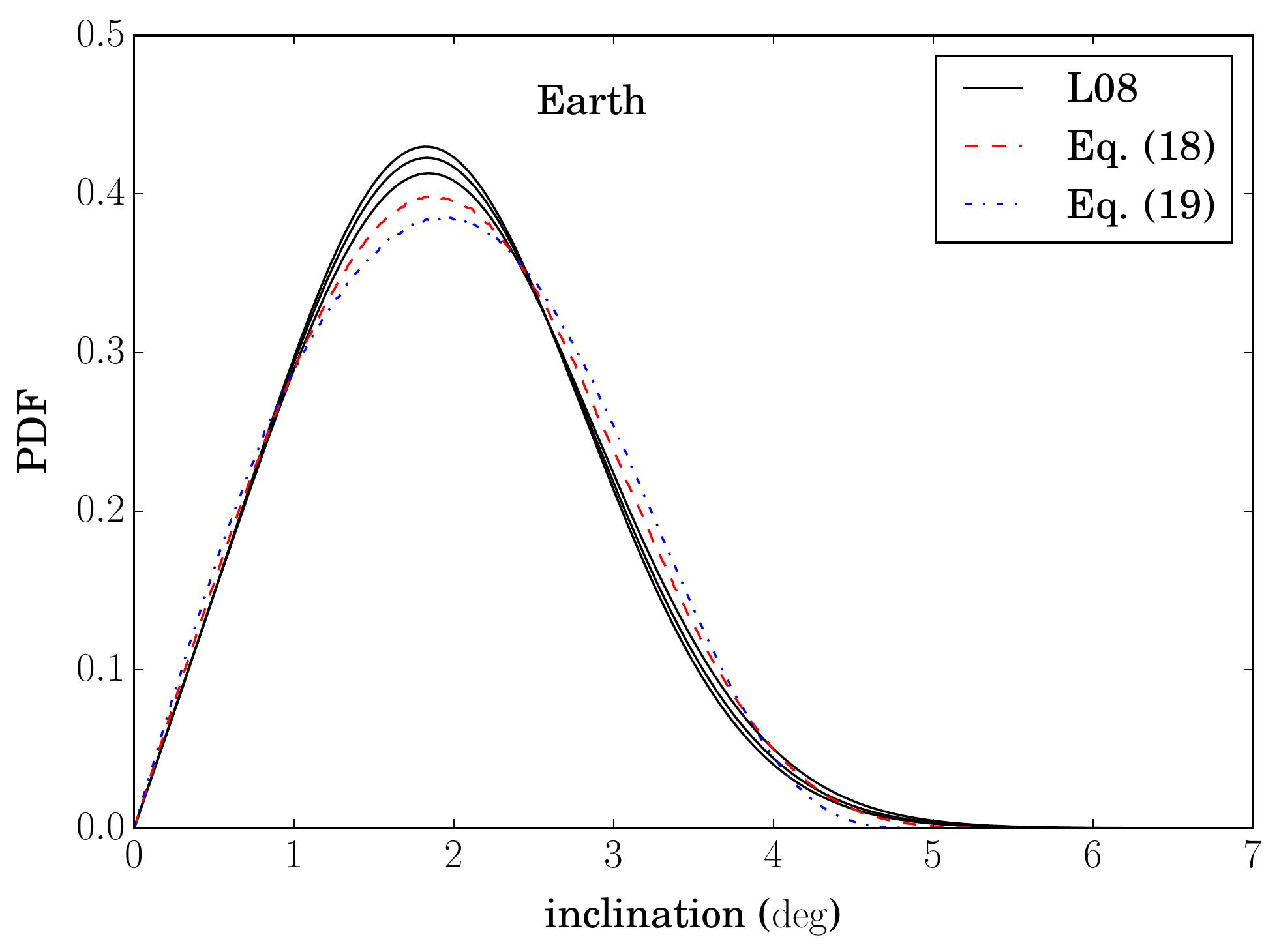}}
\caption{PDF of Earth inclination. The same as Fig.~\ref{figure3} with the addition of the prediction of Eq.~\eqref{eq:complete_ergodic_ansatz} in dash-dot line.}
\label{figure5}
\end{figure}
\begin{table}
\caption{Partition of the current AMD in the solar system. Units are AU, solar mass and yr. All quantities are multiplied by $10^{10}$.}
\label{table1}
$
\begin{array}{cccc}
\hline\hline
\noalign{\smallskip}
\mathrm{Inner\,AMD} & \mathrm{Outer\,AMD} & \sum_{k=0}^4 P'_k +Q'_k & \sum_{k=5}^8 P'_k +Q'_k  \\
\noalign{\smallskip}
\hline
\noalign{\smallskip}
567 & 357543 & 537 & 357573 \\
\noalign{\smallskip}
\hline
\end{array}
$
\end{table}
Indeed, with this choice the conservation of the total AMD reduces to that of the quantity $\sum_{k=0}^4 P'_k +Q'_k$, whose initial value is very close to the initial total AMD of the inner planets, as shown in Table~\ref{table1}. The remaining action variables, corresponding to the modes $k \in \{1, 2, 3 ,4\}$, can then be allowed to vary stochastically according to equiprobability. The statistics of the giant planet orbits would therefore be virtually identical to the one shown in Fig.~\ref{figure1}.

\begin{figure}
\centering
\resizebox{\hsize}{!}{\includegraphics{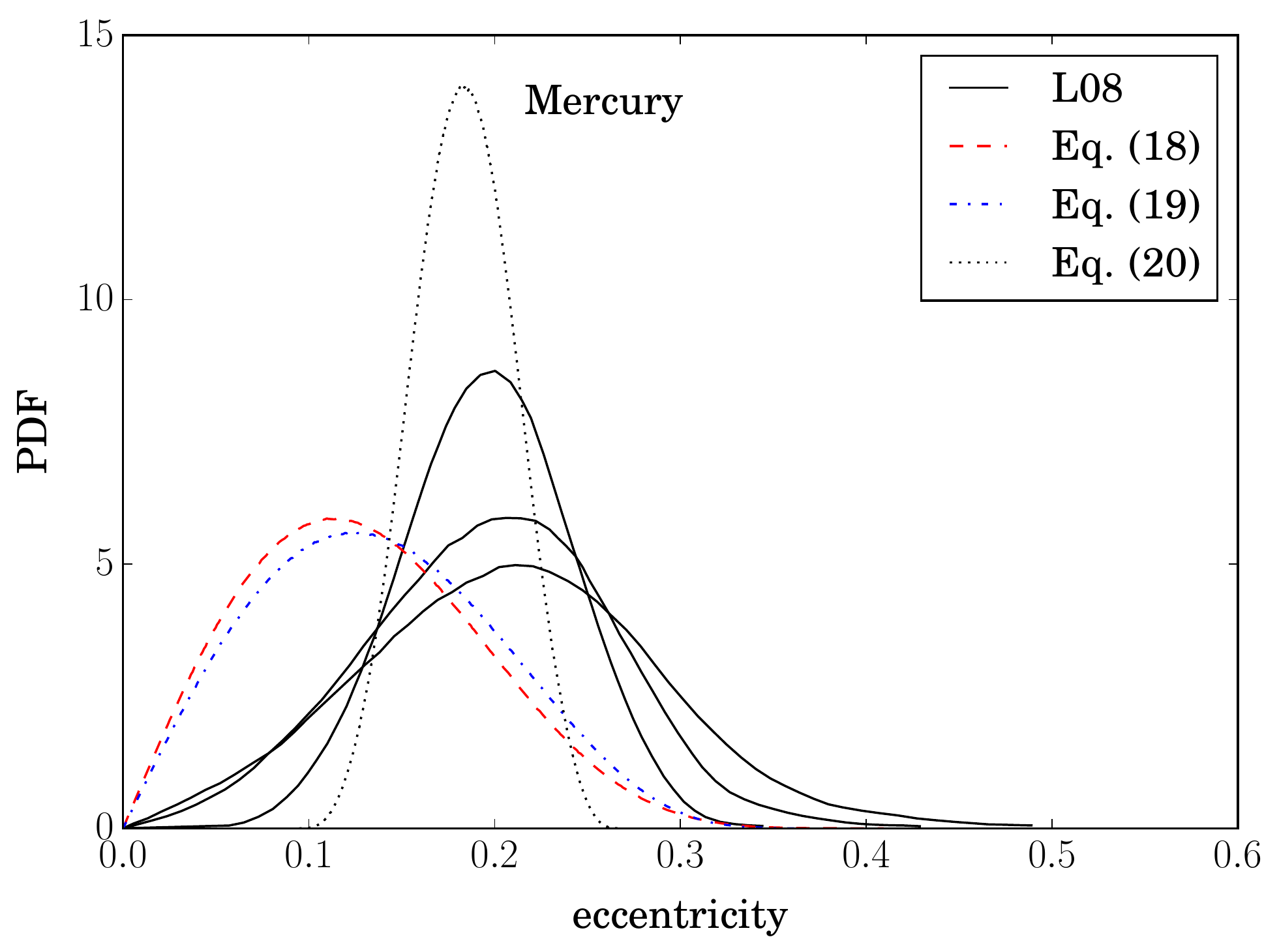}}
\caption{PDF of Mercury eccentricity. The same as Figure~\ref{figure3} with the addition of the predictions of Eq.~\eqref{eq:complete_ergodic_ansatz} in dash-dot line and Eq.~\eqref{eq:modified_ergodic_ansatz} in dotted line.}
\label{figure6}
\end{figure}

Among the integrals of motion, the AMD is of particular interest. For instance, \citet{laskar1997, laskar2000} has highlighted its role in the orbital configuration of planetary systems. An important reason for this relevance is that, differently from Eqs. \eqref{eq:quad_hamiltonian_2} and \eqref{eq:angular_momentum_Laplace_Lagrange}, the simple Eq. \eqref{eq:deficit} is valid to all orders in eccentricities and inclinations. I then start to investigate the predictions of equiprobability based on AMD conservation only. Conservation of energy will be added later. The ansatz corresponding to the above considerations reads:
\begin{equation}
\label{eq:ergodic_ansatz}
\begin{aligned}
&\rho(\vec{P'}, \vec{Q'}, \vec{p'}, \vec{q'}) \propto \delta(C - \overline{C}) \, f \\
&f(\vec{P'}, \vec{Q'}, \vec{p'}, \vec{q'}) = \delta(q'_5 - \overline{q'_{5}}) \prod_{k=5}^{8} \delta(P'_k - \overline{P'_{k}}) \, \delta(Q'_k - \overline{Q'_{k}})
\end{aligned}
\end{equation}
Conservation of $L_x$ and $L_y$ to second order in eccentricities and inclinations is contained in this distribution function as the variables $Q'_5$ and $q'_5$ are set to their initial values (see Eq. \eqref{eq:convervation_Lx_Ly}). I do not have a simple analytical formula for the PDFs of eccentricity and inclination. However, they can be calculated numerically very rapidly by direct sampling, as illustrated in the Appendix~\ref{sampling}. Moreover, I present an analytical characterization of these PDFs in the Appendix~\ref{ergodic_pdf}. The predictions of the ansatz~\eqref{eq:ergodic_ansatz} are compared in Fig.~\ref{figure3} with Laskar's PDFs. They reproduce very satisfactorily Venus and Earth PDFs. A detailed comparison in the case of Earth is shown in Figs.~\ref{figure4} and \ref{figure5}. The agreement is remarkable if one considers the simplicity of the ansatz and that it does not contain any free parameter\footnote{To fit his PDFs Laskar employs three free parameters for each planet and each orbital element (eccentricity or inclination).}. For what concerns the PDFs of Mars and that of Mercury inclination, the predictions are still reasonable as the differences with the numerical PDFs are of the same order of magnitude of the PDF diffusion over time. Indeed, in these cases one could a priori interpret the predicted curves as the long-term, stationary limit of Laskar's PDFs. However, this interpretation breaks down when one considers the eccentricity PDF of Mercury, as shown in detail in Fig.~\ref{figure6}. In this case, the predicted PDF is substantially different from the numerical one. In particular, the mean of Laskar's curve is not correctly reproduced by the ansatz. It is clear that equiprobability, as applied in Eq.~\eqref{eq:ergodic_ansatz}, is far to approximate the actual Mercury dynamics.

One can add to the ansatz~\eqref{eq:ergodic_ansatz} the conservation of secular energy at quadratic order in eccentricities and inclinations (see Eq.~\eqref{eq:quad_hamiltonian_2}),
\begin{equation}
\label{eq:complete_ergodic_ansatz}
\rho(\vec{P'}, \vec{Q'}, \vec{p'}, \vec{q'}) \propto \delta(C - \overline{C}) \, \delta(\hamLL - \overline{\hamLL}) \, f
\end{equation}
In the Appendix~\ref{sampling} I suggest an algorithm to efficiently sampling this distribution function. The addition of energy conservation does not change in a substantial way the predictions shown in Fig.~\ref{figure3}. The principal consequence is that the average AMD in a single LL mode depends now, for $k \le 4$, on the mode considered via the specific values of the frequencies $g_k$ and $s_k$. In contrast, Eq.~\eqref{eq:ergodic_ansatz} predicts an average AMD which is independent of the particular mode (see Appendix~\ref{ergodic_pdf}). This slightly shifts the predicted PDFs, without changing the validity of the above considerations. This is shown in the case of Earth in Figs.~\ref{figure4} and \ref{figure5}, and in Fig.~\ref{figure6} for Mercury.

The major disagreement between the predictions of equiprobability in the phase-space and Laskar's curves occurs for the eccentricity of Mercury, which is the less massive planet and has the highest typical eccentricity and inclination\footnote{This is a crucial point, as this study takes into account conservation of $L_x$, $L_y$ and $\hamsec$ only at quadratic order in eccentricities and inclinations (see Eqs. \eqref{eq:quad_hamiltonian_2} and \eqref{eq:convervation_Lx_Ly}).}. As suggested above, a kind of equiprobability ansatz could still be valuable if applied to a more restricted phase space domain than that defined by the conservation of the total inner planet AMD and energy. If, as a matter of speculation, one restricts the equiprobability domain by setting the action variable $P'_1$ to its initial value, i.e.
\begin{equation}
\label{eq:modified_ergodic_ansatz}
\rho(\vec{P'}, \vec{Q'}, \vec{p'}, \vec{q'}) \propto \delta(C - \overline{C}) \, \delta(\hamLL - \overline{\hamLL}) \, \delta(P'_1 - \overline{P'_1}) \, f
\end{equation}
the predicted PDF of Mercury eccentricity starts to reproduce the characteristic mean value of the corresponding Laskar's curve, as shown in Fig.~\ref{figure6}. This is related to the fact that the chaotic dynamics of Mercury eccentricity keeps a strong memory of the dominant LL mode $P'_1$\footnote{As \citet{boue2012} and \citet{batygin2015} show, the dynamics of Mercury treated as a test particle in the gravitational field of the other planets, whose orbits are fixed, can be described by a time-independent Hamiltonian, which is therefore an integral of motion and keeps memory of the initial conditions.}.
%\footnote{Therefore, according to LL dynamics, one has $P_1 \simeq P'_1$. Since $P_1$ is the eccentricity contribution to Mercury AMD and the major one to its current value, setting $P'_1$ to its initial value means freezing a relevant fraction of its AMD. In a model where Mercury is treated as a test particle and the other planets fixed on circular orbits, its AMD is indeed an integral of motion.}
However, it is clear that a much more involved analysis, taking into account high order terms in the Hamiltonian \eqref{eq:relativity}, is needed to closely reproduce the numerical PDF.

\section{Random walk approach to chaotic diffusion}
\label{random_walk}
Reproducing the PDFs of the inner planets is particularly difficult because of the time dependence of these curves. The complexity of describing diffusion of the integrals of motion in weakly nonlinear systems, starting from a given partition between linear modes, is perhaps best illustrated by the famous Fermi--Pasta--Ulam--Tsingou problem \citep{fermi1955}. As stated in Sect.~\ref{laskar}, in L08 the inner planet PDFs are fitted by means of a Rice distribution \citep{rice1945}, whose parameters depend on time to account for the curve diffusion. The choice of this particular distribution is justified by Laskar's suggestion that, because of chaotic diffusion, Poincar\'{e} variables \eqref{eq:poincare_variables} could behave like independent Gaussian random variables over a long timescale. In what follows I try, in the simplest way, to frame this ansatz in the context of the present paper, showing the difficulties of such a random walk approach. 

In the following discussion I focus on the degrees of freedom related to eccentricity, but a similar analysis applies to the inclinations ones. According to the microcanonical distribution function \eqref{eq:microcanonical_giant}, the PDF of the rectangular variable $h_k = e_k \sin(p_k)$ (i.e. marginalized over the remaining Poincar\'{e} variables) can easily be shown to be
\begin{equation}
\label{eq:x_pdf}
\rho(h_k) = \frac{1}{2 \pi} \int_{-\infty}^{+\infty} du \, e^{-i h_k u} \prod_{n=1}^{N} J_0 \left(\gamma_{kn} u \right)
\end{equation}
where $i = \sqrt{-1}$ is the imaginary unit, $J_0$ the Bessel function of the first kind and order zero, and $\gamma_{kn} = \tens{O_A}_{kn} (2\overline{P'_n}/\Lambda_k)^{1/2}$. Depending on the coefficients $\gamma_{kn}$, the PDF \eqref{eq:x_pdf} can be double-peaked and strongly different from a Gaussian distribution. Moreover, since $\rho(-h_k) = \rho(h_k)$, the mean value of $h_k$ is zero. It seems doubtful whether chaotic diffusion, acting on the LL modes, can drive Poincar\'{e} variables to the Gaussianity suggested by Laskar. Therefore, I set up the random walk approach on a slight different basis. On a secular timescale $\tsec $ of few Myr, the motion of the inner planets is well reproduced by a quasiperiodic time function. On a Lyapunov timescale, $\tLyap \simeq 5$ Myr, chaos starts to decorrelate the actual motion from the initial quasiperiodic approximation. On a much more longer timescale, $t \gg \tLyap$, the precise nature of the chaotic terms in the Hamiltonian \eqref{eq:relativity} is maybe irrelevant in assessing the general form of the inner planet PDFs. Considering such a case, I propose the following ansatz, in which the stochastic dynamics of Poincar\'{e} variables $x'_k$ and $y'_k$ is given by the superposition of a secular term and a chaotic one modelled by means of a white noise:
\begin{equation}
\label{eq:ansatz}
\vec{r} = \vec{r}_{\mathrm{sec}} + \vec{r}_{\mathrm{chaos}}
\end{equation}
where $\vec{r} = (x'_k, y'_k)$\footnote{Since the action variables $\vec{P'}$ are conserved in the LL dynamics, it is reasonable to consider them as a starting point. However, as the boundary condition $P'_k \ge 0$ applies, the most simple and straightforward approach is to use Poincar\'{e} variables $x'_k$ and $y'_k$ which are defined on the entire real line.}. The stationary random variable $\vec{r}_{\mathrm{sec}}$ describes the statistics of the LL dynamics and corresponds to the microcanonical ensemble \eqref{eq:microcanonical_giant},
\begin{equation}
\label{eq:LL_pdf}
\rho_{\mathrm{sec}}(x'_k, y'_k) = \frac{\delta \left[ r - \left(2 \overline{P'_k} \right)^{1/2} \right]}{2 \pi \left(2 \overline{P'_k} \right)^{1/2}}
\end{equation}
where $r^2 = {x'_k}^2 + {y'_k}^2$. The time-dependent chaotic component is modelled by the following Langevin equation:
\begin{equation}
\label{eq:chaos_pdf}
\vec{\dot{r}}_{\mathrm{chaos}} = \sqrt{2 D_k} \, \vec{\eta}(t)
\end{equation}
where $\vec{\eta}$ is a white noise, i.e.
\begin{equation}
\label{eq:white_noise}
\begin{aligned}
\langle \vec{\eta}(t) \rangle &= \vec{0} \\ 
\langle \eta_x(t) \eta_x(t')  \rangle &= \delta(t-t') \quad \langle \eta_y(t) \eta_y(t')  \rangle = \delta(t-t') \\
\langle \eta_x(t) \eta_y(t')  \rangle &= 0
\end{aligned}
\end{equation}
for all times $t, t' \geq 0$. For simplicity I have assumed an isotropic diffusion in the phase space of the LL mode $k$. The diffusion coefficient $D_k$ is a free parameter in this model. Moreover, I choose the initial condition on the chaotic term to match the LL dynamics,
\begin{equation}
\label{eq:r_chaos_time_0}
\vec{r}_{\mathrm{chaos}}(t=0) = \vec{0}
\end{equation}
Since the random variables $\vec{r}_{\mathrm{sec}}$ and $\vec{r}_{\mathrm{chaos}}$ are independent, the PDF of $\vec{r}$ is given by the convolution of the PDF of $\vec{r}_{\mathrm{sec}}$, $\rho_{\mathrm{sec}}$, and that of $\vec{r}_{\mathrm{chaos}}$, $\rho_{\mathrm{chaos}}$. Therefore one has
\begin{equation}
\label{eq:r_pdf}
\rho(\vec{r}, t) = \int_{\mathbb{R}^2} d\vec{z} \, \rho_{\mathrm{sec}}(\vec{z}) \, \rho_{\mathrm{chaos}}(\vec{r}-\vec{z},t)
\end{equation}
The PDF of the random variable satisfying Eqs. \eqref{eq:chaos_pdf}, \eqref{eq:white_noise} and \eqref{eq:r_chaos_time_0} is the Green function of the Fokker-Planck equation
\begin{equation}
\label{eq:fokker_planck}
\frac{\partial \rho_{\mathrm{chaos}}(\vec{\xi},t)}{\partial t} = D_k \, \vec{\nabla}^2 \rho_{\mathrm{chaos}}(\vec{\xi},t)
\end{equation}
This Green function is
\begin{equation}
\label{eq:green_function}
\rho_{\mathrm{chaos}}(\vec{\xi},t) = \frac{1}{4 \pi D_k t} \exp \left( - \frac{\vec{\xi}^2}{4 D_k t} \right)
\end{equation}
Substituting Eqs.~\eqref{eq:LL_pdf} and \eqref{eq:green_function} in Eq. \eqref{eq:r_pdf} one gets
%\begin{equation}
%\label{eq:x_y_chaotic_distribution}
%\rho(y'_k, x'_k, t) = \frac{1}{2 \pi \sigma_k^2} \exp \left( - \frac{{x'_k}^2 + {y'_k}^2 + m_k^2}{2 \sigma_k^2} \right) I_0 \left( \frac{m_k \sqrt{{x'_k}^2 + {y'_k}^2}}{\sigma_k^2} \right)
%\end{equation}
\begin{equation}
\label{eq:r_chaotic_distribution}
\rho(r, t) = \frac{r}{\sigma^2} \exp \left( - \frac{r^2 + m^2}{2 \sigma^2} \right) I_0 \left( \frac{r \, m}{\sigma^2} \right)
\end{equation}
where $I_0$ indicates the modified Bessel function of the first kind and order zero, $m = (2 \overline{P'_k})^{1/2}$ and $\sigma^2 = 2 D_k t$. Therefore, one obtains a Rice distribution for the variable $r$. In the case where the variables $x_k$ and $y_k$ are dominated by one particular LL mode, i.e. $|\gamma_{kl}| \gg |\gamma_{kn}|$ for a certain index $l$ and for all $n \neq l$, the PDF of the eccentricity $e_k$ turns out to be a Rice distribution too, with parameters $m = |\tens{O_A}_{kl}| \, (2 \overline{P'_l}/\Lambda_k)^{1/2}$ and $\sigma^2 = 2 D_k (\tens{O_A}_{kl})^2 t/\Lambda_k$. In this limiting case one thus recovers Laskar's ansatz. This is due to the fact that the Rice distribution describes the envelope of a sine wave plus Gaussian noise \citep{rice1945}, as already mentioned in L08. However, within the present framework, the parameter $m$ is not a free parameter as in L08, but it is set up by the initial LL dynamics.
\begin{table}
\caption{The parameter $m$ of a Rice distribution reproducing the PDF of the eccentricity of the inner planets. The value in L08 is compared to the one predicted by Eq.~\eqref{eq:r_chaotic_distribution} when one considers for each planet only the LL mode with the biggest amplitude, indicated by the symbol $\star$ \citep[see also][Table 4]{bretagnon1974}. $T$ is the time in Gyr.}
\label{table2}
\centering
$
\begin{array}{ccc}
\hline\hline
\noalign{\smallskip}
k & \mathrm{L08} & |\tens{O_A}_{k\star}| \, (2 \overline{P'_{\star}}/\Lambda_k)^{1/2}  \\
\noalign{\smallskip}
\hline
\noalign{\smallskip}
1& 0.1875 & 0.1810 \\
2 & 0.02235 + 0.00014 \, T & 0.0190 \\
3 & 0.01951 + 0.00013 \, T & 0.0160 \\
4 & 0.06437 - 0.00188 \, T & 0.0698 \\
\noalign{\smallskip}
\hline
\end{array}
$
\end{table}
I compare in Table \ref{table2} the value of the parameter $m$ in L08 (Table 6) for the eccentricity PDFs to the prediction of Eq.~\eqref{eq:r_chaotic_distribution} when one considers for each planet only the LL mode with the biggest amplitude. The physical meaning of the parameter $m$, which is free in L08, clearly emerges in the present framework.

As shown in Table~\ref{table2}, the PDF \eqref{eq:r_chaotic_distribution} casts some light on Laskar's fits in L08. Nevertheless, it cannot reproduce the numerical PDFs accurately, even if the diffusion coefficients $D_k$ are free parameters. There are a couple of reasons for that. Firstly, the initial distribution~\eqref{eq:LL_pdf} is based on the LL dynamics. However, terms of higher order in eccentricities and inclinations in the Hamiltonian~\eqref{eq:relativity} are of primary importance for the inner planets. To have a faithful representation of their initial quasiperiodic dynamics one has to go beyond the LL approximation.
%Indeed, to fit the numerical PDFs with a Rice distribution L08 considers a variance parameter of the form $\sigma^2 = b_0 + b_1 t$, where both $b_0$ and $b_1$ are free parameters. In the present framework, the term $b_0$ can be interpreted as the variance of an initial Gaussian distribution for the LL action variables, instead of Eq.~\eqref{eq:LL_pdf}.
The other principal limitation of the present framework is that PDF~\eqref{eq:r_chaotic_distribution} cannot reproduce the slightly decreasing peak positions of Mars PDFs. Indeed, one has $\langle P'_k \rangle = \langle r^2 \rangle/2 =  m^2/2 + \sigma^2 = \overline{P'_k} + 2 D_k t$, which means that the average content of AMD in the LL mode $k$ is an increasing function of time. Therefore, the PDFs of eccentricity and inclination can only diffuse towards increasing mean values of the corresponding random variable. The reason for this behaviour is that the random walk considered above does not take into account neither the conservation of AMD nor that of energy. Indeed, if one assumes that the AMD of the inner planets is conserved over time, it is understandable that Mercury, Earth and Venus PDFs diffuse towards increasing mean values, while Mars PDFs diffuse towards decreasing ones. Therefore, it would be a considerable improvement to set up a random walk exploiting the secular integrals of motion. Such a modification would end in a feedback dynamics between $\vec{r}_{\mathrm{sec}}$ and $\vec{r}_{\mathrm{chaos}}$ in Eq.~\eqref{eq:ansatz}.

\section{Conclusions}
\label{conclusions}
Motivated by the chaotic dynamics of solar system planets and the stochastic nature of planet formation, I address in the present paper the construction of a statistical description of planetary orbits. I suggest that such an approach can be based on the statistical mechanics of secular dynamics. Considering the solar system as a benchmark, I try to reproduce the PDFs of eccentricity and inclination calculated numerically by \citet{laskar2008}. I show that the statistics of the giant planet orbits is well reproduced by the microcanonical ensemble of the Laplace-Lagrange linear secular dynamics. This is particularly relevant as such a theory can be directly applied to a generic planetary system. Minor discrepancies in this description are connected to higher-than-quadratic terms in the planet Hamiltonian. Their contribution could be taken into account by improving the perturbation approach via successive quasi-identity canonical transformation and normal forms. I then try to reproduce the statistics of the inner planet orbits through the ansatz of equiprobability in the phase-space constrained by the secular integrals of motion, namely angular momentum and energy. Within the limitations of a stationary model, such an ansatz allows to easily and accurately reproduce the structure of Venus and Earth PDFs without any free parameter. However, major discrepancies rise in the case of Mercury eccentricity. I finally show the difficulties of constructing a random walk to model the chaotic diffusion of the inner planet PDFs, following the original ansatz of \citet{laskar2008}. Within certain limitations, the PDF I obtain allows to illustrate the physical meaning of one of the free parameters in Laskar's ansatz.

It is clear that the above description of the inner planet statistics has to be improved. One has to taken into account higher-than-quadratic terms in the planet Hamiltonian. Generally speaking, one could think of using the full expression of the secular energy, Eq.~\eqref{eq:relativity}. The problem would then be how efficiently sample the corresponding microcanonical ensemble. Another viable approach could be to start constructing an ad hoc model for Mercury, since it shows the most relevant discrepancies with respect to the proposed ansatz. Such an approach could consist in considering Mercury as a test particle in the gravitational field of the other solar system planets, whose orbits are fixed \citep{boue2012, batygin2015}. With a selection of the most important higher order terms relevant for the non-linear dynamics of Mercury, important improvements could be achieved in reproducing the PDFs of its eccentricity and inclination. With respect to the random walk approach presented in Sect.~\ref{random_walk}, a considerable improvement would be taking into account the conservation of angular momentum deficit and energy. Even if such a model is not completely predictive, as the diffusion in the phase space is described by free parameters, it could be nevertheless really instructive in judging how the conservation of the integrals of motion constrains the relaxation of the Laplace-Lagrange action variables.

The relevance of a statistical description of planet orbits in a generic planetary system is particularly manifest if one considers the model of planet spacing by \citet{laskar2000} \citep[see also][]{laskar2017}. In this approach to planet formation, Laskar describes the collisionless dynamics of planetary embryos by a stochastic variation in their orbital elements that conserves the total angular momentum deficit. Conservation of mass and linear momentum is taken into account to model the result of collisions, during which the total angular momentum deficit decreases. Collisions stop when the total angular momentum deficit is too smalls to allow for further close encounters. With these simple assumptions, Laskar is able to show how the structure of a generic planetary system derives from the initial mass distribution of planetary embryos. As already pointed out by \citet{tremaine}, Laskar's model is not unique because it requires an ad hoc prescription for the random evolution of the orbital elements between collisions. Statistical descriptions like those in Sects.~\ref{giant} and \ref{inner} can provide such a prescription based on solid physical arguments. A statistical approach to planetary orbits could then be useful in studying planet formation, in combination with standard N-body simulation of planetary trajectories.

%-------------------------------------------------------------------
\begin{acknowledgements}
I am grateful to A. Morbidelli, J. Touma, J. Laskar, J. Teyssandier, N. Cornuault and L. Pittau for the fruitful discussions. I would also like to thank J.-P. Beaulieu for his support and help in reviewing the manuscript.
\end{acknowledgements}

%-------------------------------------------------------------------
\bibliographystyle{bibtex/aa.bst}    % style aa.bst
\bibliography{bibtex/statistical_mechanics.bib}    % References file.

\begin{appendix}

\section{General relativistic correction}
\label{secular_hamiltonian}
I present a short derivation of the leading general relativistic contribution to the secular planetary Hamiltonian. At the order $(v/c)^2$ the Lagrangian of Sun and planets is given by \citep{landau1975}
\begin{equation}
\label{eq:Lagrangian}
\begin{split}
L &= L_0 + 3 \sum_a {\sum_b}^\prime \frac{m_a v^2_a}{2} \frac{G m_b}{c^2 r_{ab}}
+ \sum_a \frac{m_a v^4_a}{8 c^2} \\
&- \sum_a {\sum_b}^\prime \frac{G m_a m_b}{4 c^2 r_{ab}} \left[ 7 \, \vec{v}_a \cdot \vec{v}_b + (\vec{v}_a \cdot \vec{n}_{ab})( \vec{v}_b \cdot \vec{n}_{ab})  \right] \\
&- \sum_a {\sum_b}^\prime {\sum_c}^\prime \frac{G^2 m_a m_b m_c}{2 c^2 r_{ab} r_{ac}}
\end{split}
\end{equation}
where $L_0 = \sum_a m_a v^2_a/2 + \sum_a {\sum}^\prime_b G m_a m_b/2 r_{ab}$ is the Newtonian Lagrangian. In the equation above $a, b, c \in \{0, 1, \dots, 8\}$, $r_{ab} = |\vec{r}_a - \vec{r}_b|$, $n_{ab} = (\vec{r}_a - \vec{r}_b)/r_{ab}$ and the prime symbol means that the terms $b = a$ and $c = a$ must be excluded from the summation. I also recall that $m_i/m_0 \ll 1$ for $i \in \{1, 2, \dots, 8\}$. I then simplify Eq.~\eqref{eq:Lagrangian} by keeping among the relativistic terms only those of order $(v/c)^2$. I thus neglect terms of order $(m_i/m_0)(v/c)^2$ and smaller. One obtains $L = L_0 + \delta L$, with
\begin{equation}
\label{eq:delta_Lagrangian}
\delta L = \sum_i \frac{m_i}{2 c^2} \left( \frac{v^4_i}{4} + 3\frac{G m_0 v^2_i }{r_i} - \frac{G^2 m^2_0}{r^2_i} \right)
\end{equation}
where $r_i$ is the Heliocentric position of planet $i$. The general relativistic correction to the planetary Hamiltonian is thus given by $\delta \ham = - \delta L$ \citep{landau1969}. To obtain the secular correction one has to average $\delta \ham$ over the Keplerian orbits of the non-interacting planets. Neglecting corrections of the order $m_i/m_0$, along a Keplerian orbit one has
\begin{equation}
\label{eq:Keplerian_velocity}
v^2_i = \frac{G m_0}{a_i} \left( \frac{2 a_i}{r_i} - 1 \right)
\end{equation}
where $a_i$ is the semi-major axis of planet $i$. Moreover, averaging over an orbital period one obtains $\langle r_i^{-1} \rangle = a_i^{-1}$ and $\langle r_i^{-2} \rangle = a_i^{-2} (1-e^2_i)^{-1/2}$, with $e_i$ the eccentricity of planet $i$. Substituting Eq.~\eqref{eq:Keplerian_velocity} in $\delta \ham$ and taking the average, one finally finds
\begin{equation}
\label{eq:delta_Hamiltonian_average}
\langle \delta \ham \rangle = \sum_i \frac{G^2 m^2_0 m_i}{c^2 a^2_i} \left[ \frac{15}{8} - \frac{3}{(1-e^2_i)^{1/2}}  \right]
\end{equation}
Discarding the first term as it is constant in the secular dynamics, one obtains that the dominant secular contribution of general relativity is $\hamsecGR = - 3 G^2 m^2_0 m_i / c^2 a^2_i (1-e^2_i)^{1/2}$.

\section{Laplace-Lagrange Hamiltonian}
\label{matrices}
The matrix $\tens{A}$ appearing in the quadratic Hamiltonian \eqref{eq:quad_hamiltonian} is given by the following expression:
\begin{equation}
\label{eq:A}
\tens{A}_{kl} = \left\{\begin{array}{@{}l@{\quad}l}
      - \frac{G m_k m_l a_k a_l}{\pi \sqrt{\Lambda_k \Lambda_l} |a_k - a_l|^3} I_{2kl} & \mbox{if $k \neq l$} \\[\jot]
      \sum_{\substack{j=1 \\ j \neq k}}^{N} \frac{G m_k m_j a_k a_j}{\pi \Lambda_k |a_k - a_j|^3} I_{1kj} + \frac{3 G^2 m_0^2 m_k}{c^2 a_k^2} & \mbox{if $ k = l$}
    \end{array}\right.
\end{equation}
%\begin{equation}
%\label{eq:A}
%\tens{A}_{kl} =
%\begin{dcases}
%- \frac{G m_k m_l a_k a_l}{\pi \sqrt{\Lambda_k \Lambda_l} |a_k - a_l|^3} I_{2kl} & \mbox{if } k \neq l \\
%\sum_{\substack{j=1 \\ j \neq k}}^{N} \frac{G m_k m_j a_k a_j}{\pi \Lambda_k |a_k - a_j|^3} I_{1kj} + \frac{3 G^2 m_0^2 m_k}{c^2 a_k^2} & \mbox{if } k = l
%\end{dcases} 
%\end{equation}
where I have defined
\begin{equation*}
I_{nkl} = \int_{0}^{\frac{\pi}{2}} d\theta \frac{\cos(2n \theta)}{\left[ 1 + 4 a_k a_l \sin^2 \theta / (a_k - a_l)^2 \right]^{\frac{3}{2}}}.
\end{equation*}
Similarly, the matrix $\tens{B}$ is given by
\begin{equation}
\label{eq:B}
\tens{B}_{kl} = \left\{\begin{array}{@{}l@{\quad}l}
      \frac{G m_k m_l a_k a_l}{\pi \sqrt{\Lambda_k \Lambda_l} |a_k - a_l|^3} I_{1kl} & \mbox{if $k \neq l$} \\[\jot]
      - \sum_{\substack{j=1 \\ j \neq k}}^{N} \frac{G m_k m_j a_k a_j}{\pi \Lambda_k |a_k - a_j|^3} I_{1kj} & \mbox{if $ k = l$}
    \end{array}\right.
\end{equation}
%\begin{equation}
%\label{eq:B}
%\tens{B}_{kl} =
%\begin{dcases}
%\frac{G m_k m_l a_k a_l}{\pi \sqrt{\Lambda_k \Lambda_l} |a_k - a_l|^3} I_{1kl} & \mbox{if } k \neq l \\
%- \sum_{\substack{j=1 \\ j \neq k}}^{N} \frac{G m_k m_j a_k a_j}{\pi \Lambda_k |a_k - a_j|^3} I_{1kj} & \mbox{if } k = l
%\end{dcases} 
%\end{equation}
From formula \eqref{eq:B} it is easy to check that matrix $\tens{B}$ verifies the following relation:
\begin{equation}
\label{eq:B_zero_eigenvalue}
\sum_{l=1}^{N} \tens{B}_{kl} \sqrt{\Lambda_l} = 0
\end{equation}
This implies that one of the $s_k$ eigenvalues is zero and $\left. \left(\! \! \sqrt{\Lambda_1}, \dots, \! \! \sqrt{\Lambda_N} \right) \, \right/ \! \! \sqrt{\sum_k\Lambda_k}$ is, up to a sign, the corresponding normalized eigenvector.

\section{Ansatz of equiprobability: analytical description}
\label{ergodic_pdf}
An analytical characterization of the PDFs of eccentricity and inclination predicted by Eq.~\eqref{eq:ergodic_ansatz} can be obtained by writing the conservation of the total AMD of the inner planets, as motivated in Sect.~\ref{inner}, through the modified Delauney variables~\eqref{eq:delaneuy}. Compared to that of Sect.~\ref{inner}, such an approach does not include the statistics of giant planet orbits. The microcanonical density of states $\omega$ then reads
\begin{equation}
\label{eq:microcanonical_density}
\frac{\omega(C_0)}{(2\pi)^{2N}} = \left[ \prod_{k=1}^{N} \int_0^{\Lambda_k} \! \! \! dP_k\! \! \int_0^{2(\Lambda_k-P_k)} \! \! \! dQ_k \right] \delta \left( C_0 - \sum_{k=1}^N P_k + Q_k \right)
\end{equation}
where $N=4$ in the present analysis, $C_0$ is the initial total AMD of the inner planets and the factor $(2\pi)^{2N}$ comes from the integration over the angle variables $p_k$ and $q_k$. Since $\Lambda_k \geq 3C_0/2$ for all $k$ \citep[Table 1]{laskar1997}, one can restrict the above integration to the hypercube $[0,C_0]^{2N}$.
%\begin{equation}
%\label{eq:microcanonical_density_2}
%\frac{\omega(C_0)}{(2\pi)^{2N}} = \left[ \prod_{k=1}^{N} \int_0^{C_0} \! \! \! dP_k\! \! \int_0^{C_0} \! \! \! dQ_k \right] \, \, \delta \left( C_0 - \sum_{k=1}^N P_k + Q_k \right)
%\end{equation}
To calculate the density of states one can therefore write $\omega(C_0) = d \Sigma(C_0)/dC_0$, with
\begin{equation}
\label{eq:Sigma}
\Sigma(C_0) = \left[ \prod_{k=1}^{N} \int_0^{C_0} \! \! \! dP_k\! \! \int_0^{C_0} \! \! \! dQ_k \right] \, \, H \left( C_0 - \sum_{k=1}^N P_k + Q_k \right)
\end{equation}
where $H$ is the Heaviside step function. Performing the integrations by iteration, one obtains $\Sigma(C_0) = (2\pi)^{2N} {C_0}^{2N}/2N!$ and then
\begin{equation}
\label{eq:microcanonical_density_3}
\omega(C_0) = (2\pi)^{2N} \frac{{C_0}^{2N-1}}{(2N-1)!}
\end{equation}
The joint microcanonical PDF of $P_k$ and $Q_k$ (i.e. marginalized over the remaining modified Delauney variables) is then given by
\begin{equation}
\label{eq:rho_P_Q}
\rho(P_k, Q_k) = \frac{(2\pi)^{2N}}{\omega(C_0)} \frac{d}{dC_0} \left[ \frac{(C_0 - P_k - Q_k)^{2N-2}}{(2N-2)!} \right]
\end{equation}
for $P_k + Q_k \leq C_0$, otherwise it is zero. Thus one obtains
\begin{equation}
\label{eq:rho_P_Q_2}
\rho(P_k, Q_k) = \frac{(2N-1)(2N-2)}{{C_0}^2} \left( 1 - \frac{P_k + Q_k}{C_0} \right)^{2N-3}
\end{equation}
for $P_k + Q_k \leq C_0$. Switching to the variables $e_k$ and $i_k$, one finds
\begin{equation}
\label{eq:rho_e_i}
\begin{split}
\rho(e_k, i_k) = &\frac{(2N-1)(2N-2)}{\eta^2} e_k \sin i_k \,  \cdot \\
&\cdot \left( 1 - \frac{1 - \sqrt{1-e_k^2} \cos i_k}{\eta} \right)^{2N-3}
\end{split}
\end{equation}
for $\eta^{-1} \left(1 - \sqrt{1-e_k^2} \cos i_k \right) \leq 1$, with $\eta = C_0/\Lambda_k$. The PDF of $e_k$ is then given by
\begin{equation}
\label{eq:rho_ecc}
\rho(e_k) = \int_0^{i^{\star}} di_k \,\, \rho(e_k, i_k), \quad e_k \leq \sqrt{1-(1-\eta)^2}
\end{equation}
where $\cos i^\star = (1-\eta)/(1-e_k^2)^{1/2}$. Performing the integral, one obtains
\begin{equation}
\label{eq:rho_ecc_2}
\rho(e_k) = \frac{2N-1}{\eta} \frac{e_k}{\sqrt{1-e_k^2}} \left[ 1 - \eta^{-1} \left(1 - \sqrt{1-e_k^2} \right) \right]^{2N-2}
\end{equation}
for $e_k \leq \sqrt{1-(1-\eta)^2}$. Similarly, to obtain the PDF of $i_k$ one has to calculate
\begin{equation}
\label{eq:rho_inc}
\rho(i_k) = \int_0^{e^{\star}} de_k \,\, \rho(e_k, i_k), \quad \cos i_k \geq 1-\eta
\end{equation}
where $e^\star = [1 - (1-\eta)^2/\cos^2i_k]^{1/2}$. Performing the integration, one finds
\begin{equation}
\label{eq:rho_inc_2}
\begin{split}
\rho(i_k) = &\tan i_k \, \left[ 1 - \eta^{-1}(1-\cos i_k) \right]^{2N-2} \, \cdot \\
&\cdot \left[ \frac{2N-1}{\eta}- \frac{1 - \eta^{-1}(1-\cos i_k)}{\cos i_k} \right]
\end{split}
\end{equation}
for $\cos i_k \geq 1-\eta$. As in the present study one has $\eta \ll 1$ \citep[Table 1]{laskar1997}, one can simplify these PDFs obtaining
\begin{align}
\label{eq:rho_ecc_final}
\rho(e_k) &= \frac{2N-1}{\eta} e_k \left( 1 - \frac{e_k^2}{2\eta} \right)^{2N-2}, \quad e_k \leq \sqrt{2\eta} \\
\label{eq:rho_inc_final}
\rho(i_k) &= \frac{2N-1}{\eta} i_k \left( 1 - \frac{i_k^2}{2\eta} \right)^{2N-2}, \quad i_k \leq \sqrt{2\eta}
\end{align}
In this limit one has that
\begin{align}
\label{eq:averages}
\langle e_k \rangle &= \langle i_k \rangle =  \sqrt{\frac{\pi}{2}}\frac{\Gamma(2N)}{\Gamma(2N+1/2)} \eta^{1/2} \\
\langle e^2_k \rangle &= \langle i^2_k \rangle = \frac{\eta}{N}
\end{align}
where $\Gamma$ is the Gamma function, and the average AMD of planet $k$ is thus given by
\begin{equation}
\label{eq:AMD_average}
\langle P_k + Q_k \rangle = \Lambda_k \frac{\langle e_k^2 + i_k^2 \rangle}{2} = \frac{C_0}{N}
\end{equation}
Therefore one obtains the equipartition of AMD between planets and between the eccentricity and inclination degrees of freedom. Even though they do not coincide, the PDFs~\eqref{eq:rho_ecc_final} and \eqref{eq:rho_inc_final} present the same overall structure as those predicted by the ansatz~\eqref{eq:ergodic_ansatz}.

One can also include in the above analytical treatment the conservation of the angular momentum components $L_x$ and $L_y$ at quadratic order in eccentricities and inclinations (see Eq.~\eqref{eq:angular_momentum_Laplace_Lagrange}). While the shape of the eccentricity PDF is unaffected, that of the inclination PDF turns out to be somewhat influenced by the parameter $(L_x^2 + L_y^2)^{1/2}/L_z$.

\section{Ansatz of equiprobability: sampling}
\label{sampling}
Conservation of AMD and energy in Eqs.~\eqref{eq:ergodic_ansatz} and \eqref{eq:complete_ergodic_ansatz} is equivalent to the two following conditions:
\begin{align}
\label{eq:AMD_energy_conservation}
C_0 &= \sum_{k=1}^{4} P'_k + Q'_k  \\
H_0 &= - \sum_{k=1}^{4} \left( g_k P'_k + s_k Q'_k \right)
\end{align}
with $C_0 = \sum_{k=1}^{4} \overline{P'_k} + \overline{Q'_k}$ and $H_0 = - \sum_{k=1}^{4} \left( g_k \overline{P'_k} + s_k \overline{Q'_k} \right)$. These equations can be rewritten as
\begin{align}
1 &= \sum_{k=1}^{4} a_k + b_k  \label{sampling1} \\
h &= \sum_{k=1}^{4} g_k a_k + s_k b_k \label{sampling2}
\end{align}
where $a_k = P'_k/C_0$, $b_k = Q'_k/C_0$ and $h = - H_0/C_0$. Moreover, I recall that $a_k, b_k \geq 0$, $g_k > 0$ and $s_k < 0$ .  The PDF \eqref{eq:ergodic_ansatz} can be therefore evaluated by an uniform sampling of $a_k$ and $b_k$ verifying Eq.~\eqref{sampling1}. To evaluate the PDF~\eqref{eq:complete_ergodic_ansatz} $a_k$ and $b_k$ have to verify both Eq.~\eqref{sampling1} and Eq.~\eqref{sampling2}.

Uniform sampling of $N$ positive real numbers that add up to unity, as in Eq.~\eqref{sampling1}, can be performed via a direct sampling algorithm. One starts by sampling $N-1$ random numbers uniformly from the interval $[0,1]$. After sorting them, one obtains the set $\{u_1, \dots, u_{N-1}\}$, with $u_{i-1} \leq u_{i}$ for $i \in \{2,\dots,N-1\}$. Then, the numbers $\{u_1, u_2-u_1, \dots, u_{N-1}-u_{N-2}, 1-u_{N-1} \}$ follow the target distribution.

To uniformly sampling positive real numbers $a_k$ and $b_k$ from Eqs.~\eqref{sampling1} and \eqref{sampling2} there is probably no direct sampling method. However, I present here an efficient algorithm with rejection. Let's assume that $h \geq 0$, as in the present study. A similar algorithm can be constructed in case $h$ is negative. One can multiply Eq.~\eqref{sampling1} by $g_{\mathrm{max}} = \max_k\{a_k\}$, subtract Eq.~\eqref{sampling2} and finally divide by $g_{\mathrm{max}}-h$ (one can assume this to be different from zero, otherwise the sampling is trivial). One obtains
\begin{equation}
\label{eq:new_sampling}
1 = \sum_{k=1}^4 \frac{g_{\mathrm{max}}-g_k}{g_{\mathrm{max}}-h} a_k + \frac{g_{\mathrm{max}}-s_k}{g_{\mathrm{max}}-h} b_k
\end{equation}
The coefficients of $a_k$ and $b_k$ in Eq.~\eqref{eq:new_sampling} are positive and the term $a_{k^\star}$, such that $g_{k^\star} = g_{\mathrm{max}}$, does not appear in it. Therefore, the other $N-1$ numbers can be sampled from this equation using the direct sampling algorithm presented above. If their sum $S$ turns out to be smaller or equal to unity, then one can take $a_{k^\star} = 1 - S$ to obtain a set of numbers following the target distribution. Otherwise, the numbers are rejected and Eq.~\eqref{eq:new_sampling} sampled again.

\end{appendix}

\end{document}